\PassOptionsToPackage{unicode}{hyperref}
\PassOptionsToPackage{hyphens}{url}
\documentclass[
]{article}
\usepackage{lmodern}
\usepackage{amssymb,amsmath}
\usepackage{ifxetex,ifluatex}
\ifnum 0\ifxetex 1\fi\ifluatex 1\fi=0 
  \usepackage[T1]{fontenc}
  \usepackage[utf8]{inputenc}
  \usepackage{textcomp} 
\else 
  \usepackage{unicode-math}
  \defaultfontfeatures{Scale=MatchLowercase}
  \defaultfontfeatures[\rmfamily]{Ligatures=TeX,Scale=1}
\fi
\IfFileExists{upquote.sty}{\usepackage{upquote}}{}
\IfFileExists{microtype.sty}{
  \usepackage[]{microtype}
  \UseMicrotypeSet[protrusion]{basicmath} 
}{}
\makeatletter
\@ifundefined{KOMAClassName}{
  \IfFileExists{parskip.sty}{%
    \usepackage{parskip}
  }{
    \setlength{\parindent}{0pt}
    \setlength{\parskip}{6pt plus 2pt minus 1pt}}
}{
  \KOMAoptions{parskip=half}}
\makeatother
\usepackage{xcolor}
\IfFileExists{xurl.sty}{\usepackage{xurl}}{} 
\IfFileExists{bookmark.sty}{\usepackage{bookmark}}{\usepackage{hyperref}}
\hypersetup{
  pdftitle={A Realistic Guide to Making Data Available Alongside Code to Improve Reproducibility.},
  pdfauthor={Nicholas J Tierney* (1,2) \& Karthik Ram* (3)},
  hidelinks,
  pdfcreator={LaTeX via pandoc}}
\usepackage[margin=1in]{geometry}
\usepackage{longtable,booktabs}
\usepackage{etoolbox}
\makeatletter
\patchcmd\longtable{\par}{\if@noskipsec\mbox{}\fi\par}{}{}
\makeatother
\IfFileExists{footnotehyper.sty}{\usepackage{footnotehyper}}{\usepackage{footnote}}
\makesavenoteenv{longtable}
\usepackage{graphicx}
\makeatletter
\def\maxwidth{\ifdim\Gin@nat@width>\linewidth\linewidth\else\Gin@nat@width\fi}
\def\maxheight{\ifdim\Gin@nat@height>\textheight\textheight\else\Gin@nat@height\fi}
\makeatother
\setkeys{Gin}{width=\maxwidth,height=\maxheight,keepaspectratio}
\makeatletter
\def\fps@figure{htbp}
\makeatother
\providecommand{\tightlist}{%
  \setlength{\itemsep}{0pt}\setlength{\parskip}{0pt}}
\usepackage{float}
\let\origfigure\figure
\let\endorigfigure\endfigure
\renewenvironment{figure}[1][2] {
    \expandafter\origfigure\expandafter[H]
} {
    \endorigfigure
}

\RequirePackage{setspace}
\spacing{1.5}

\RequirePackage{fancyhdr}
\fancypagestyle{plain}{%
\fancyhf{} 
\fancyfoot[C]{\sffamily\thepage} 

}

\RequirePackage{microtype}

\RequirePackage{caption}
\DeclareCaptionStyle{italic}[justification=centering]
 {labelfont={bf},textfont={it},labelsep=colon}
\usepackage[utf8]{inputenc}
\usepackage{booktabs}
\usepackage{longtable}
\usepackage{array}
\usepackage{multirow}
\usepackage{wrapfig}
\usepackage{float}
\usepackage{colortbl}
\usepackage{pdflscape}
\usepackage{tabu}
\usepackage{threeparttable}
\usepackage{threeparttablex}
\usepackage[normalem]{ulem}
\usepackage{makecell}
\usepackage{xcolor}
\newlength{\cslhangindent}
\newenvironment{cslreferences}%
  {\setlength{\parindent}{0pt}%
  \everypar{\setlength{\hangindent}{\cslhangindent}}\ignorespaces}%
  {\par}

\title{A Realistic Guide to Making Data Available Alongside Code to Improve Reproducibility.}
\author{Nicholas J Tierney* (1,2) \& Karthik Ram* (3)}
\date{}

\begin{document}
\maketitle

1 = Monash University, Department of Econometrics and Business Statistics, Melbourne, Australia

2 = Australian Centre of Excellence for Mathematical and Statistical Frontiers (ACEMS)

3 = Berkeley Institute for Data Science, University of California, Berkeley, USA

* = Both authors contributed equally to the work

Correspondence to:

Nicholas Tierney (\href{mailto:nicholas.tierney@gmail.com}{\nolinkurl{nicholas.tierney@gmail.com}}) \& Karthik Ram (\href{mailto:k@karthik.io}{\nolinkurl{k@karthik.io}})

\hypertarget{abstract}{%
\section*{Abstract}\label{abstract}}
\addcontentsline{toc}{section}{Abstract}

Data makes science possible. Sharing data improves visibility, and makes the research process transparent. This increases trust in the work, and allows for independent reproduction of results. However, a large proportion of data from published research is often only available to the original authors. Despite the obvious benefits of sharing data, and scientists' advocating for the importance of sharing data, most advice on sharing data discusses its broader benefits, rather than the practical considerations of sharing. This paper provides practical, actionable advice on how to actually share data alongside research. The key message is sharing data falls on a continuum, and entering it should come with minimal barriers.

\hypertarget{intro}{%
\section{Introduction}\label{intro}}

\begin{quote}
``Data! data! data!'' he cried impatiently. ``I can't make bricks without clay.'' - Sherlock Holmes (The Adventure of the Copper Beeches by Sir Arthur Conan Doyle)
\end{quote}

Data are a fundamental currency upon which scientific discoveries are made. Without access to good data, it becomes extremely difficult, if not impossible, to advance science. Yet, a large majority of data on which published research papers are based rarely see the light of day and are only visible to the original authors (Rowhani-Farid and Barnett 2016; Stodden, Seiler, and Ma 2018). Sharing data sets upon publication of a research paper has benefits to individual researchers, often through increased visibility of the work (Popkin 2019; Kirk and Norton 2019). But there is also a component of benefit to the broader scientific community. This primarily comes in the form of potential for reuse in other contexts along use for training and teaching (McKiernan et al. (2016)). Assuming that the data have no privacy concerns (e.g., human subjects, locations of critically endangered species), or that the act of sharing does not put the authors at a competitive disadvantage (data can be embargoed for reasonable periods of time), sharing data will always have a net positive benefit. First and foremost, sharing data along with other artifacts can make it easier for a reader to independently reproduce the results, thereby increasing transparency and trust in the work. The act of easy data sharing can also improves model training, as many different models can be tried and tested on latest data sources, closing the loop on research and application of statistical techniques. Existing data sets can be combined or linked with new or existing data, fostering the development and synthesis of new ideas and research areas. The biggest of these benefits is the overall increase in reproducibility.

For nearly two decades, researchers who work in areas related to computational science have pushed for better standards to verify scientific claims, especially in areas where a full replication of the study would be prohibitively expensive. To meet these minimal standards, there must be easy access to the data, models, and code. Among the different areas with a computational bent, the bioinformatics community in particular has a strong culture around open source (Gentleman et al. 2004), and has made releasing code and associated software a recognized mainstream activity. Many journals in these fields have also pushed authors to submit code (in the form of model implementations) alongside their papers, with a few journals going as far as providing a ``reproducibility review'' (Peng 2011).

\textbf{In this paper we focus on the practical side of sharing data for the purpose of reproducibility.} Our goal is to describe various methods in which an analyst can share their data with minimal friction. We steer clear of idealistic ideas such as the FAIR data principles (Wilkinson et al. 2016) since they still do not help a researcher share their data. We also skip the discussion around citation and credit because data citations are still poorly tracked and there is no standardized metric or a h-index equivalent for data as of this writing.

For a piece of computational research to be minimally reproducible, it requires three distinct elements: 1) Code; 2) Computing environment, and 3) Data. The first two of these challenges have largely been solved (Poisot et al. 2019).

Although code sharing in science had a rocky start (Barnes 2010), more and more code writing by scientists is being shared, partly due to the rapid increase in training opportunities made available by organizations like The Carpentries, combined with the rapid adoption of Github by scientists (Ram 2013). The bigger driver for this may also be connected to the rise in popularity of data science as a discipline distinct from statistics (Donoho 2017). This rapid growth in data science has largely been fueled by easy access to open source software tools. Programming languages such as Python, R and Julia help scientists implement and share new methods to work with data (R Core Team 2019; ``Python,'' n.d.; Bezanson et al. 2017). Each of these languages is supported by thriving communities of researchers and software developers who contribute many of the building blocks that make them popular. As of this writing, Python, Julia, and R have 167k packages (``Pypi,'' n.d.), \textasciitilde{} 14k packages (``Cran,'' n.d.) and \textasciitilde2k packages (``Julia-Pkgman,'' n.d.) respectively. These packages form the building blocks of a data scientists daily workflow.

In a modern computational workflow a user can \texttt{pip\ install} a package in Python, or use \texttt{install.packages} in R to install all software dependencies for a project. By relying on the idea of having research compendia (Gentleman and Temple Lang 2007) or something as simple as a requirements file, it is relatively straightforward to install all the necessary scaffolding. Data on the other hand are rarely accessible that easily.

A typical data analysis loads a dozen or two of these open source libraries at the top of a notebook and then relies on existing routines to rapidly read, transform, visualize, and model data. Each package depends on a complex web of other packages, building upon existing work rather than re-implementing everything from scratch. Working from script and a list of such dependencies, a data analyst can easily install all the necessary tools in any local or remote environment and reproduce the computation. When new functionality is developed, it is packaged into a separate entity and added to a language's package manager.

The computing environment is also easily captured with modern tools such as Docker (Boettiger 2015; Jupyter 2018). Modern tools such as Binder (Jupyter 2018) can parse Docker files and dependency trees to provide on demand, live notebooks in R and Python that a reader can immediately execute in the browser without dealing with the challenges of local installation. This makes it simple to load a specific environment to run any analysis. Code is handled by version control with tools like Git and GitHub (``Git,'' n.d.; ``Github,'' n.d.), paired with archiving frameworks such as Zenodo provide access to code (particularly model implementations)(Zenodo 2016). All the necessary software is available from various package managers (and their numerous geographic mirrors and archives) making it easy to install any version of a software package. However, the biggest challenge, even today, remains easy and repeatable access to data in a data analysis.

Datasets are often far more diverse than code in terms of complexity, size, and formats. This makes them particularly challenging to standardize or easily ``install'' where the code is running. While there are numerous public and private data repositories, none of them function as package managers, which is what provides so much robustness to code sharing. As a result, data used in an analysis is often read from various locations (local, network, or cloud), various formats, varying levels of tidiness (Wickham 2014). There is also the overhead associated with data publishing, the act of archiving data in a repository that also mints permanent identifiers, that are not required of all projects due to the effort and time involved. It is worth drawing the distinction between data sharing (making the data available with little effort) and data publishing (archiving the data in a long-term repository, with our without curated metadata).

What aspects of data make them particularly challenging to share from a reproducibility perspective? This is the question we tackle in this paper. While there are several papers that serve as best-practice guides for formatting data (Broman and Woo 2017) and getting them ready for sharing (Ellis and Leek 2017), the aims of this paper are a bit different. Our aim is to address the issue of data in the context of reproducibility in data science workflows. In particular we discuss the various tradeoffs one has to consider when documenting and sharing data, when it is worth publishing, and how this would change depending on the use case, perceived impact, and potential audience.

We discuss how to share and/or publish data and cover various tradeoffs when deciding how much to do. We skip detailed discussions of data preparation (Broman and Woo 2017; Ellis and Leek 2017), or citation (as of this writing, data citation is still in its infancy). We also analyze the state of data contained inside software packages, shared and made available as part of modern data analysis. How often are data shipped inside software packages? How does data format and size impact availability?

\hypertarget{minimal}{%
\section{A minimal structure to share data for analysis}\label{minimal}}

To share data analysis data, there should be some minimal set of requirements. For example, the data should contain information on metadata, data dictionaries, the README, and data used in analysis (See Section \ref{minimal}). No matter where data is submitted, there should ideally be a canonical data repository in one long term archive that links to others. It should also have an accession number, or Digital Object Identification (DOI) number, which allows for it to be cited (discussed further in \ref{cite}).

There are 8 pieces of content to consider for data sharing:

\begin{enumerate}
\def\labelenumi{\arabic{enumi}.}
\tightlist
\item
  README: A Human readable description of the data
\item
  Data dictionary: Human readable dictionary of data contents
\item
  License: How to use and share the data
\item
  Citation: How you want your data to be cited
\item
  Machine readable meta data: Make your data searchable
\item
  Raw data: The original/first data provided
\item
  Scripts: To clean raw data ready for analysis
\item
  Analysis ready data: Final data used in analysis
\end{enumerate}

One basic suggested directory layout is given below in Figure \ref{fig:code-directory}.

\begin{figure}

{\centering \includegraphics[width=0.75\linewidth]{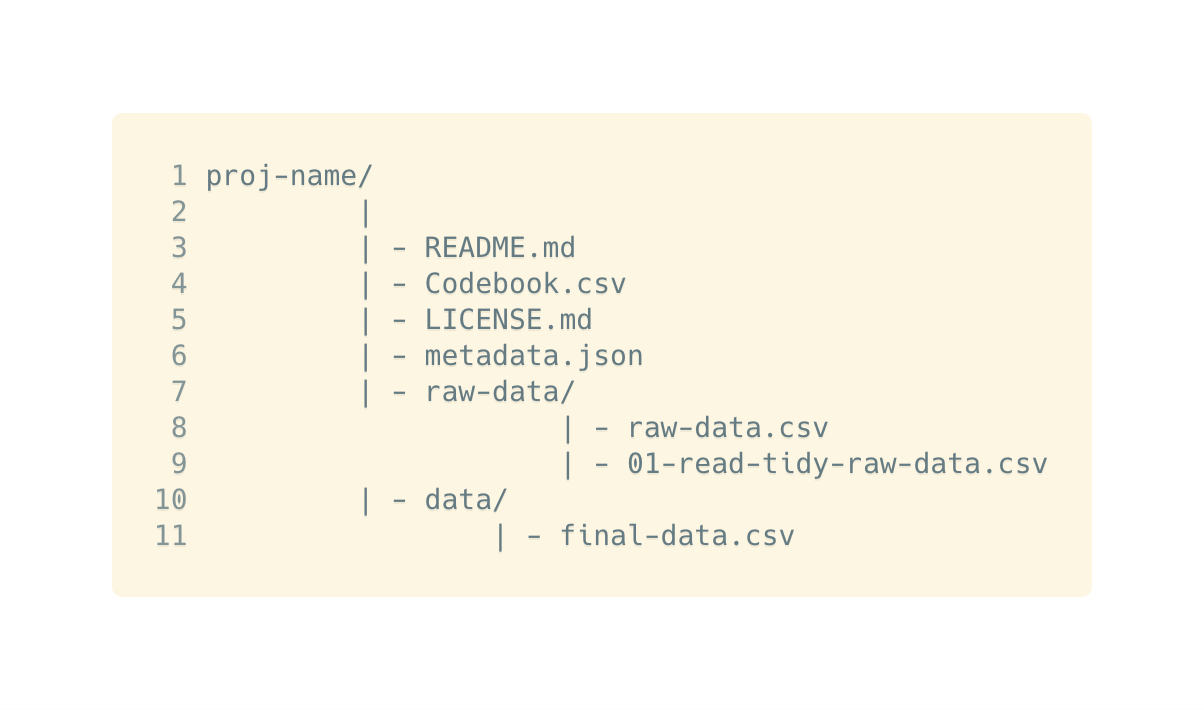} 

}

\caption{Example directory layout and structure for a data repository.}\label{fig:code-directory}
\end{figure}

From these sections, the minimal files to provide in order of most importance are:

\begin{enumerate}
\def\labelenumi{\arabic{enumi}.}
\tightlist
\item
  README
\item
  Data dictionary
\item
  License
\item
  Citation
\item
  Analysis ready data
\item
  Scripts to tidy analysis raw data into ready data
\end{enumerate}

These sections are now described.

\hypertarget{readme}{%
\subsection{README: A Human readable description of the data}\label{readme}}

In the context of datasets, READMEs are particularly useful when there are no reliable standards. A README is often the first place people will go to learn more about anything in a folder - they are very common in software, and historically were included so the uppercase letters of README meant it would be at the top of a directory. The README is meant for someone to read and understand more about the data and contains the ``who, what, when, where, why, \& how'':

\begin{itemize}
\tightlist
\item
  \textbf{Who} collected it
\item
  \textbf{What} is the data
\item
  \textbf{When} was it collected
\item
  \textbf{Where} was it collected
\item
  \textbf{Why} it was collected
\item
  \textbf{How} is was collected
\end{itemize}

The README should be placed in the top level of the project, and with one README per dataset. It should be brief, and provide links to the other aforementioned sections, and be in one directory. It should also contain any other guidance for the user on how to read and interpret the whole directory. For example, explaining where tidy and raw data are stored, and other scripts for tidying.

Saving a README with the extension \texttt{.md} file gives the author the formatting benefits of \texttt{markdown}, making it simple to insert links, tables, and make lists. In systems like GitHub, a README file is detected and rendered in nice HTML on the repository by default.

\hypertarget{data-dictionary}{%
\subsection{Data dictionary: Human readable dictionary of data contents}\label{data-dictionary}}

A data dictionary provide human readable description of the data, providing context on the nature and structure of the data. This helps someone not familiar with the data understand, and use the data. At a minimum they should contain the following pieces of information about the data:

\begin{itemize}
\tightlist
\item
  \textbf{variable names}
\item
  \textbf{variable labels}
\item
  \textbf{variable codes}, and
\item
  special values for \textbf{missing data}.
\end{itemize}

\begin{figure}
\includegraphics[width=1\linewidth]{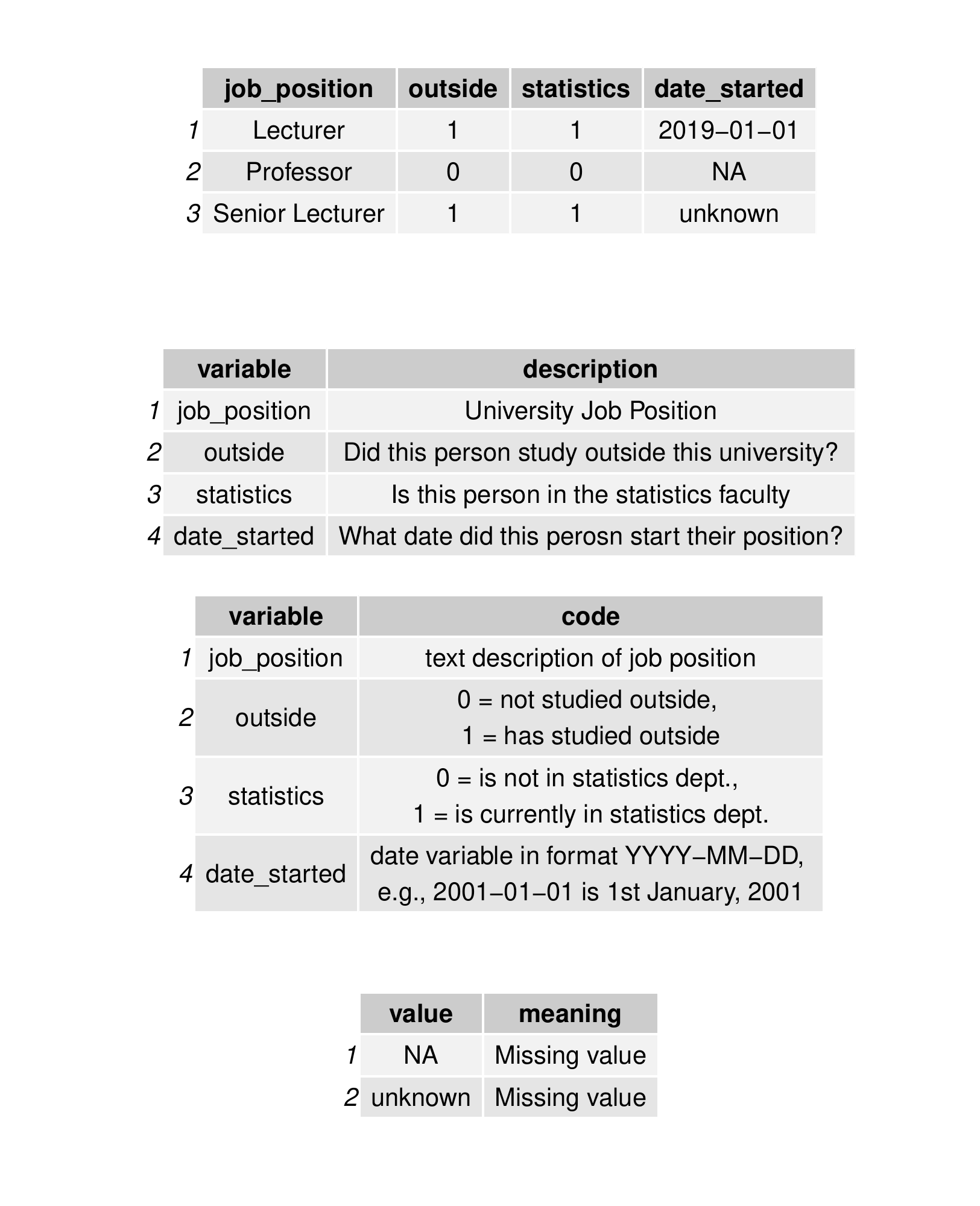} \caption{Four tables - the data, and it's variable names, variable codes, and the meaning of missing values}\label{fig:variables}
\end{figure}

Figure \ref{fig:variables} shows an example of the data and data dictionary details.
\textbf{Variable names} should be short, descriptive names with no spaces or special characters. In the most common tabular data case, these correspond to column names, for example, ``job\_position'', ``faculty\_level'', and ``years\_at\_uni''. \textbf{Variable labels} are longer descriptions of variables. For example ``University Job Position'', ``University Faculty Position Level'', and ``Number of Years Worked at University'' (``Codebook Cookbook: A Guide to Writing a Good Codebook for Data Analysis Projects in Medicine'' n.d.; Arslan 2019). \textbf{Variable codes} apply to categorical (factor) variables, and are the values for their contents. For example, a variable could contain answers to a question, with values and codes like so: \texttt{0\ =\ no,\ 1\ =\ yes}, to indicate if someone is in statistics, for example. These should be consistent across similar variables to avoid problems where \texttt{0\ =\ yes} for one variable, and \texttt{1\ =\ yes} in another. Date variables should have consistent formatting. For example, all date information could be in format ``YYYY-MM-DD'', and this should not be mixed with ``YYYY-DD-MM''. Date formatting should be described in the variable labels. \textbf{Missing data} are values that should have been observed, but were not. The code for missingness should be documented in the data dictionary, and should nominally be \texttt{NA}. If the reason for missingness is known it should be recorded. For example, censored data, patient drop out, or measurement error can have different values, such as ``unknown'', -99, or other value codes (White et al. 2013; Broman and Woo 2017).

Table \ref{tab:table-data-incarcerate} shows an example data dictionary table from the Tidy Tuesday repository on incarceration trends (``Tidy-Tuesday-Incarcerate,'' n.d.). This includes information on the variable, its class (type), and a longer description.

\begin{table}[!h]

\caption{\label{tab:table-data-incarcerate}The prisoner summary data dictionary, with columns for the variable, its class, and a short description of the contents of the variable.}
\centering
\begin{tabular}[t]{lll}
\toprule
Variable & Class & Description\\
\midrule
year & integer (date) & Year\\
urbanicity & character & County-type (urban, suburban, small/mid, rural)\\
pop\_category & character & Category for population - either race, gender, or Total\\
rate\_per\_100000 & double & Rate within a category for prison population per 100,000 people\\
\bottomrule
\end{tabular}
\end{table}

Data dictionaries should be placed in the README and presented as a table. Every data dictionary should also be provided in its raw form (e.g., a CSV) in the repository, so they aren't ``trapped'' in the README.

\hypertarget{license}{%
\subsection{License: How to use and share the data}\label{license}}

Data with a license clearly establishes rules on how everyone can modify, use, and share data. Without a license, these rules are unclear, and can lead to problems with attribution and citation. It can be overwhelming to try and find the right license for a use case. Two licenses that are well suited for data sharing are:

\begin{enumerate}
\def\labelenumi{\arabic{enumi}.}
\tightlist
\item
  Creative Commons Attribution 4.0 International Public License (CC BY), and
\item
  Creative Commons CC0 1.0 Universal (CC0)
\end{enumerate}

Other licenses or notices to be aware of are \textbf{copyrighted data}, and \textbf{data embargoes}. If you are working with data \textbf{already copyrighted}, (for example under CC BY or CC0), you must give follow appropriate guidelines for giving credit. Data may also be under \textbf{embargo}. This means data cannot be shared more widely until a specific release time. If sharing data under an embargo, include detailed information on the embargo requirements in: the README, and in separate correspondence with those who receive the data. Databases may also be licensed with the Open Data Commons Open Database License: \url{https://opendatacommons.org/licenses/odbl/}, which provides provisions for sharing, creating, and adapting, provided that work is attributed, shared, and kept open.

Once you decide on a license, you should provide a LICENSE or LICENSE.md file that contains the entire license in the top level of the directory.

\hypertarget{cc-by}{%
\subsubsection{CC BY}\label{cc-by}}

The CC BY enforces attribution and due credit by default, but gives a lot of freedom for its use. Data can be shared and adapted, even for commercial use, with the following conditions:

\begin{itemize}
\tightlist
\item
  You must provide appropriate credit to the source. This means listing the names of the creators.
\item
  Link back to the CC BY license, and
\item
  Clearly show if changes were made.
\item
  Data cannot be sub-licensed, that is - a change to the existing license
\item
  There is also no warranty, so the person or people who obtained the data cannot be held liable.
\end{itemize}

The journal PLOS Comp Bio requires that data submitted cannot be more restrictive than CC BY (``PLOS Computational Biology,'' n.d.). For a brief overview of the CC BY, suitable to include in a README, see (``CCBY Short Guide,'' n.d.). For the full license, see (``CCBY Full License,'' n.d.).

\hypertarget{cc0}{%
\subsubsection{CC0}\label{cc0}}

The CC0 is a ``public domain'' license. Data with a CC0 license means the data owners waive all their rights to the work, and it now ``owned'' by the public. The data can be freely shared, copied, modified, and distributed, even for commercial purposes \emph{without asking permission}. When using data with CC0, it is good practice to cite the original paper, but it is not required. If you wish to use the CC0, see (``Choose-Cc0,'' n.d.). For a brief overview of the CC0, see (``CC0 Short Guide,'' n.d.), and for the full license, see (``CC0 Full License,'' n.d.).

\hypertarget{cite}{%
\subsection{Citation: How you want your data to be cited}\label{cite}}

A Digital Object Identifier (DOI), is a unique identifier for a digital object such as a paper, poster, or software, and is a prerequisite for citation. For example, if the data are deposited in repositories like Dryad, Zenodo, or the Open Science Framework, the best practice would be to copy the citation created by these repositories. Under the hood, the DOI is ``minted'' by DataCite, who provide DOIs for research outputs and data. This means that when citing data, it only makes sense to cite datasets that have been deposited into a DataCite compliant repository. If a DOI is unavailable, a citation will be meaningless, as it cannot be tracked by any means. A file named \texttt{citation} should be placed in the directory, at the same level as the \texttt{README}. It should contain a DOI, and could be in \texttt{.bibtex} format.

\hypertarget{machine-metadata}{%
\subsection{Machine readable metadata}\label{machine-metadata}}

\begin{figure}

{\centering \includegraphics[width=0.75\linewidth]{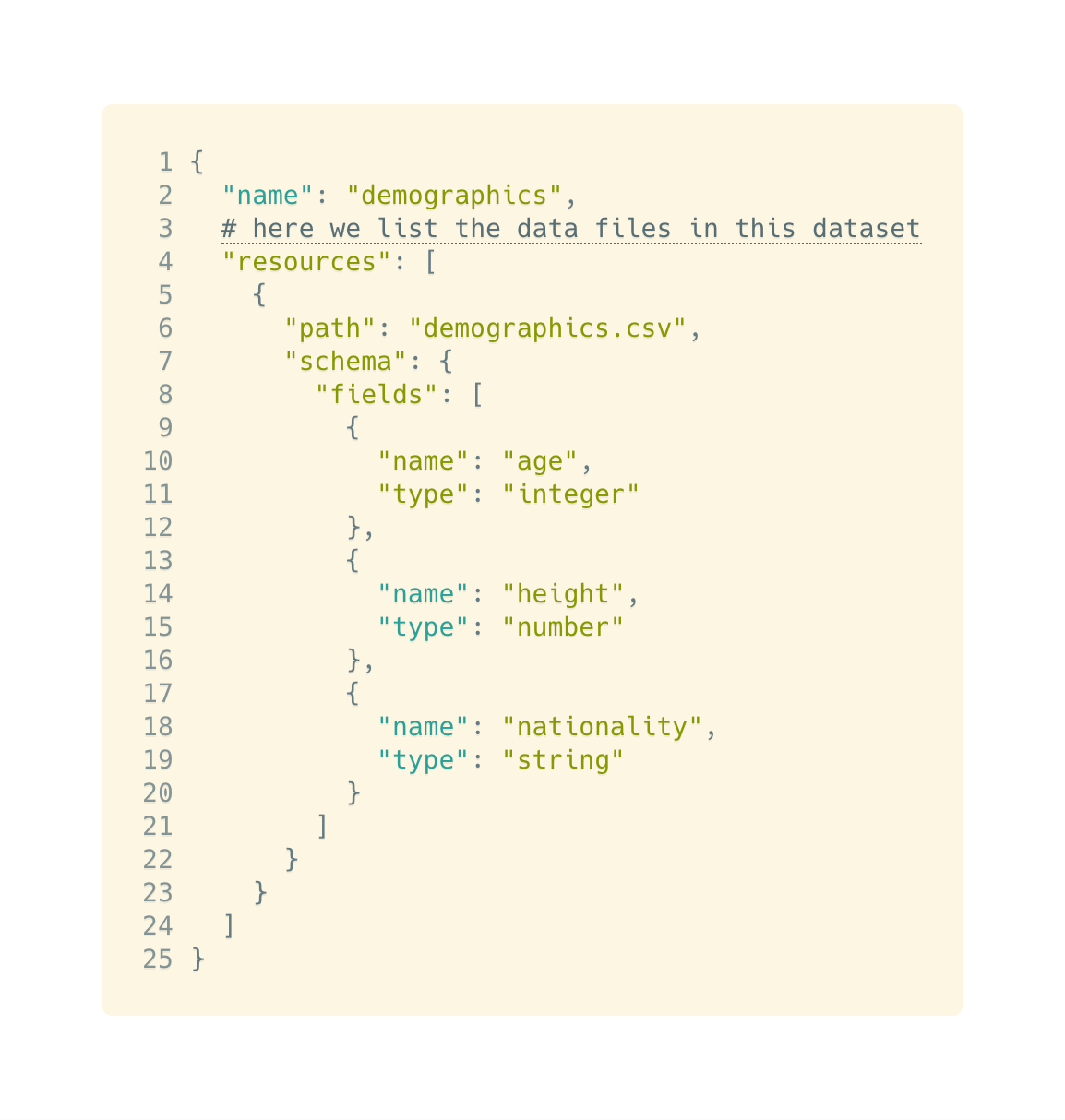} 

}

\caption{Example snippet of some Table Schema data for a dataset with three variables. This provides a description of each field, and the type of field.}\label{fig:json-code}
\end{figure}

The README and data dictionary provides \emph{human readable} information on the data. To ensure data types are preserved - for example, dates are dates, names are characters - there needs to be some form of \emph{machine readable} metadata. This provides a structure allowing the data to be indexed and searched online, through services such as google datasets search (Castelvecchi 2018). An excellent standard for metadata is Table Schema written by frictionless data (Fowler, Barratt, and Walsh 2017). For example, a dataset ``demographics'' with three variables is shown in Table \ref{tab:table-machine-readable}, and the Java Script Object Notation (JSON) equivalent in Figure \ref{fig:json-code}.

\begin{table}[!h]

\caption{\label{tab:table-machine-readable}Example demographics table of age, height, and nationality.}
\centering
\begin{tabular}[t]{rrl}
\toprule
age & height & nationality\\
\midrule
12 & 161.5 & Australian\\
21 & 181.2 & American\\
37 & 178.3 & New Zealand\\
\bottomrule
\end{tabular}
\end{table}

This contains fields such as ``path'', describing the file path to the data, and a schema with sub-fields name and type, for each variable. It also provides information for licensing and features such as line breaks, and delimiters, useful to ensure the data is correctly imported into a machine. It is built on JSON, a lightweight, human-friendly, machine readable data-interchange format. Table schema is baked into formats such as csvy (``Csvy,'' n.d.), an extended \texttt{csv} format, which has additional front matter in a lightweight markup format (YAML) using Table Schema.

In contrast to a CSV data dictionary, JSON-LD has a defined, nested structure, which means more can be communicated efficiently, while still maintaining readability, and avoiding extra writing and repetition that comes with a plain CSV where everything is defined from scratch.

Another rich format is XML, the e\textbf{X}tensible \textbf{M}arkup \textbf{L}anguage. This was an early iteration on the idea of a plain text format that was both human and machine readable. It is powerful and extensible (it powers the entire Microsoft Office suite), but for metadata purposes it is not as human readable as JSON. JSON is shorter, and quicker and easier to read and write than XML, and can also be parsed with standard JavaScript, whereas XML must be parsed by a special XML parser. XML and JSON can be converted into each others respective formats, for example with the \texttt{json} or \texttt{xml2} R packages. XML is still used as metadata for data storage, for example in EML, the Ecological Metadata Language (n.d.).

To create appropriate metadata, we recommend metadata generators such as \texttt{dataspice} or \texttt{codebook} (\emph{Dataspice: Create Lightweight Schema.org Descriptions of Dataset} 2018; Arslan 2019). These help automatically create the structured JSON data. This metadata should be provided in a folder called ``metadata'', which should be provided for every dataset.

\hypertarget{raw-data}{%
\subsection{Raw data: The original or first data provided}\label{raw-data}}

Raw data is usually the first format of the data provided before any tidying or cleaning of the data. If the raw data is a practical size to share, it should be shared in a folder called \texttt{data-raw}. The raw data should be in the form that was first received, even if it is in binary or some proprietary format. If possible, data dictionaries of the raw data should be provided in this folder as well.

\hypertarget{scripts}{%
\subsection{Scripts: To clean raw data ready for analysis}\label{scripts}}

Any code used to clean and tidy the raw data, should be provided in the \texttt{data-raw} directory. This cleaned up data, analysis ready data should be placed in the \texttt{data/} directory. Ideally this would involve only scripted languages, but if other practical steps were taken to clean up the data, these should be recorded in a plain text or markdown file.

\hypertarget{final-data}{%
\subsection{Analysis ready data: Final data used in analysis}\label{final-data}}

The data used in the data analysis should be provided in a folder called \texttt{data}. Ideally, the data ``analysis ready data'' should be in ``Tidy Data'' format (Wickham 2014), where tidy data contains variables in columns, observations in rows, and only one value per cell.

In contrast to ``raw data'', ``analysis ready data'' should be in an easily readable plain-text format, such as comma-, tab- or semicolon-separated files. These typically have file extensions like \texttt{.csv}, \texttt{.tsv} or \texttt{.txt}. Binary or proprietary formats are discouraged in favor of interoperability, as it requires special software to read, even if it is sometimes slower to read in and out, and harder to share due to size. For example, R data formats like \texttt{.rds}, \texttt{.rda}, or SPSS, STATA, or SAS data formats, \texttt{.sav}, \texttt{.dta}, or \texttt{.sas7bdat}.

\hypertarget{data-release-r}{%
\section{Releasing data in R}\label{data-release-r}}

One low cost and easy way to distribute data alongside compute is to package the datasets as a data only package or as part of something larger where the methods are being implemented. The R ecosystem has many noteworthy examples of data-only packages. One exemplar is the \texttt{nycflights13} package by Hadley Wickham (Wickham 2018), This package makes available airline data for all flights departing New York city in 2013 in a tidy format, with distinct tables for metadata on airlines, airports, weather, and planes. The package not only provides ready to use binary data but also ships raw data (in a folder called \texttt{data-raw}) along with scripts used to clean them. The package was originally created as a way to provide example data to teach tidy data principles and serves as a great model for how one can publish a data package in R.

A more common use case is to include data as part of a regular package where analytical methods are being implemented and shared. This serves the dual purpose of exposing methods and data together, making it extremely easy for a researcher to simply install and load the library at the top of their script. CRAN's distributed network (along with the global content distribution network maintained by RStudio) ensure that the package is quickly accessible to anyone in the R ecosystem. A second critical advantage in this approach is that one could also benefit from R's package documentation syntax to generate useful metadata for fast reference. This approach can also easily be adapted to other languages. Python for example, is far more flexible with respect to including arbitrary files as part of a package.

\textbf{Other benefits of packaging data in R}

\begin{enumerate}
\def\labelenumi{\arabic{enumi}.}
\item
  Packaging data can be a very powerful pedagogical tool to help researchers understand how to transform data and prepare it for further analysis. To do so, one can package raw data alongside scripts. Long form documentation such as vignettes can provide further detailed discussion on the process. Users can also skip the raw data and scripts and proceed directly to the fast binary data, which can hold a lot of data when heavily compressed.
\item
  When the primary motivation for shipping data is to illustrate visualization techniques or to run examples, one can skip the raw data and processing scripts and only include the binary data. As long as the total package size does not exceed 5 megabytes, it would be acceptable as part of CRAN. For cases when this size is hard to maintain, CRAN recommends data-only packages that will be rarely updated. For a detailed discussion on this issue and alternative approaches, see Brooke Anderson and Eddelbuettel (2017).
\end{enumerate}

Unlike CRAN, bioconductor does not have a 5 megabyte limit for package size and offers a more liberal data inclusion policy. They have a strict specification for how to organize genomic data in a package, so they can be used for data analysis out of the box. While a robust solution, it only works for homogeneous data as found with bioinformatics. For example, data in their ExperimentData section (bioconductor.org/packages/3.9/data/experiment) can be used reliably. Such a strict standard would be impossible to enforce or scale on a repository as heterogeneous as CRAN. However, CRAN packages are still generally used for demonstration or testing purposes, over generating new knowledge for papers.

One major disadvantage of packaging data inside R is that it makes the data availability very language centric. Non R users are unlikely to download and export data out of a package. This is why we recommend, as a rule, that researchers also archive data in a long-term data repository. These include domain specific repositories (see Section \ref{publish-repos}) or more general purpose ones such as Zenodo or Figshare and include the persistent identifier in all locations where the data is referenced such as the manuscript, notebook and data package.

Of the 15539 packages on Central R Archive Network (CRAN), 6278 contain
datasets either as binary data (5903 packages) or as external datasets
(766). Binary files comprise a bulk of the data shipped in the \texttt{data}
folder (68.06\%) with other plain text formats such as txt, CSV, dat,
\texttt{json} comprising less than one percent of data shipped in packages.

\hypertarget{med-large}{%
\section{Dealing with medium to large data}\label{med-large}}

Another common situation that researchers face is in dealing with data that fall somewhere between small and large. For example, small data could be tabular, as a CSV, in the order of bytes to megabytes to gigabytes that can be easily compressed, and large could be what falls under the umbrella of \textbf{big data}. The happy medium is generally data that are too big to fit on memory of most standard machines, but can successfully fit on disk (\url{https://ropensci.github.io/arkdb/articles/articles/noapi.html}). In this case, users who do not have the support to maintain resource intensive database servers can instead rely on light serverless databases such as MonetDB or SQLite. These databases provide disk based storage and using language agnostic interfaces, a analyst can easily query these data in manageable chunks that don't overrun memory limitations. Using software packages such as \texttt{arkdb} (Boettiger 2018) one could easily chunk large data from flat text files to these lite databases without running into memory limitations. Another option is to break large text data files into chunks with named sequences (e.g., \texttt{teaching-1.csv}, \texttt{teaching-2.csv}, etc.).

To make these files available alongside compute, one ingenious but short-term solution is to use the GitHub release feature to attach such large database dumps. GitHub releases are designed to serve software releases and provide the opportunity to attach binary files. This mechanism can also be used to attach arbitrary files such as large data files, binary data, and database dumps as long as each file does not exceed 2gb. The R package \texttt{piggyback} (\texttt{https://docs.ropensci.org/piggyback/}) allows for uploading and downloading such files to GitHub releases, making it easy for anyone to access data files wherever the script is being run. We emphasize again that this is a short-term solution that is dependent on GitHub maintaining this service.

\hypertarget{doco-challenge}{%
\section{Challenges in documenting your dataset}\label{doco-challenge}}

\begin{figure}
\includegraphics[width=0.9\linewidth]{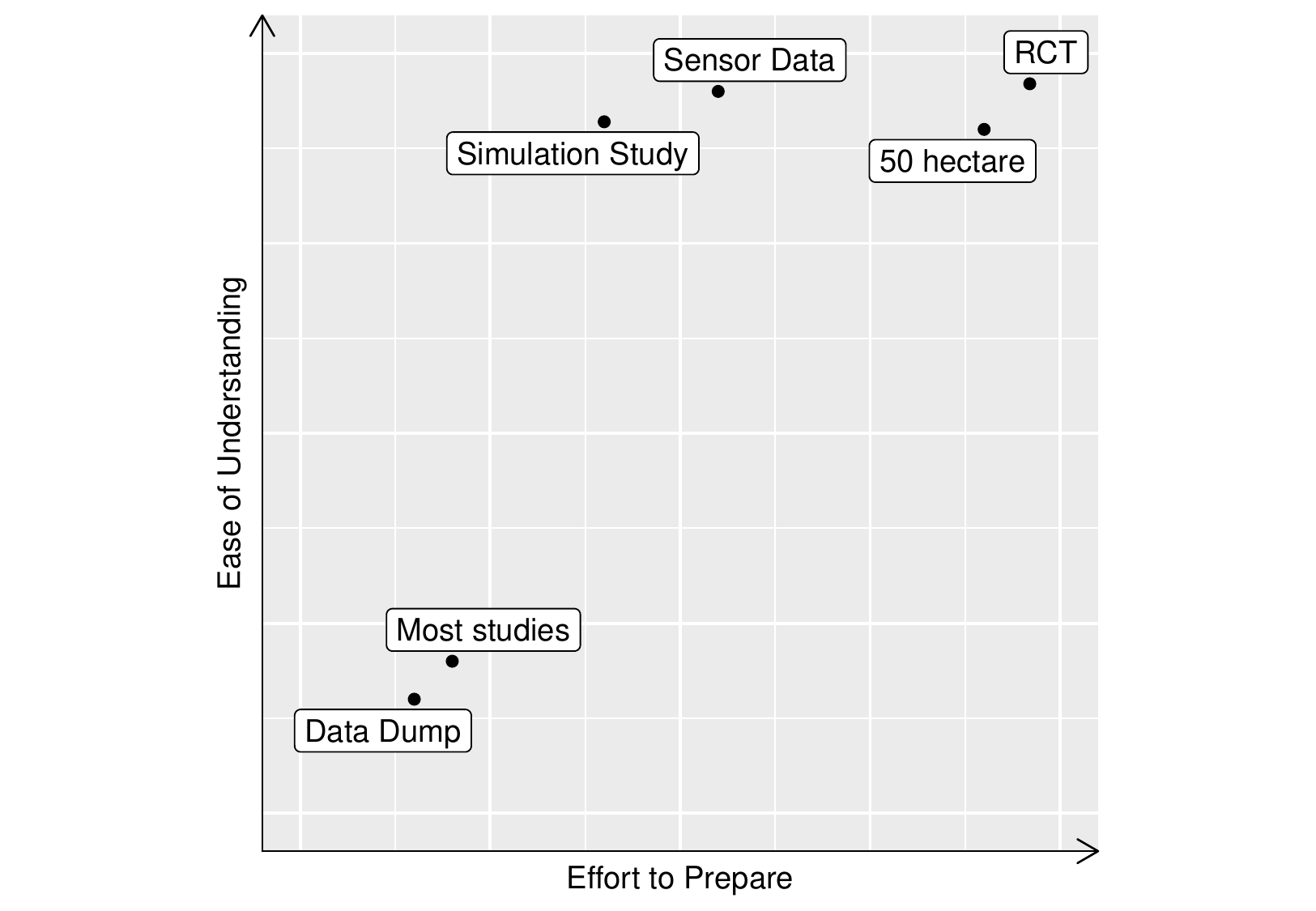} \caption{There is a big difference in the effort to prepare data, and how easy it is to understand - look at the difference between most datasets, and something like a Randomized Control Trial (RCT).}\label{fig:effort-understanding}
\end{figure}

There are many features to include with data alongside publications. However, not all of these are needed for every dataset. Working out which are needed is a challenge that is not discussed in the literature. This section discusses a framework for users, to help them decide how much they should document their data. To frame discussion around the challenges of data documenting, we can think of how an individual dataset falls on two features: ``Effort to prepare'', and ``Ease of understanding'' in Figure \ref{fig:effort-understanding}. The ideal space to be in the graph would be the top left hand corner. But what we notice is that taking more effort to prepare data means that the data is easier to understand.

Data with higher potential impact and reuse should be made as easy to understand as possible; but it also requires more time and effort to prepare. Impact is hard to measure, and varies from field to field, but as an example, take some data from medical randomized control trials (RCTs) on cancer treatment. These can have high impact, so requires a lot of time and effort to document. Comparatively, a small survey on a few demographics can have low impact. Placing these on the graph above, we see they might not have a worthwhile tradeoff for ease of understanding and ease of preparation. This means the effort put into preparing data documentation for a small survey should be kept relatively simple, not over complicated. Likewise, data that can be created via simulation from open source software could arguably not be shared since it can be generated from scratch with code; a reproducible process that requires computer time, not person time to create.

\begin{figure}
\includegraphics[width=0.9\linewidth]{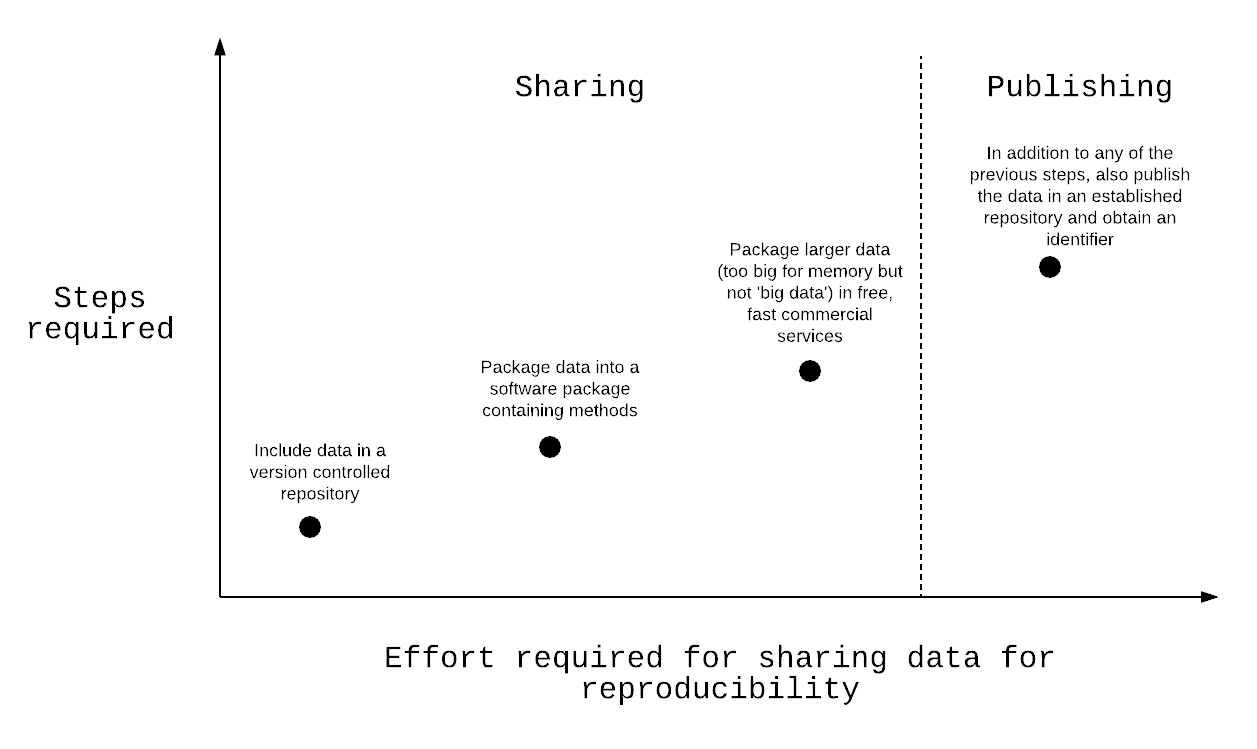} \caption{The steps required compared to effort required for data sharing for reproducibility.}\label{fig:data-sharing-workflow}
\end{figure}

Deciding how much data documentation to include should be based on the impact of the data. The more impactful the data, the more documentation features to include. Figure \ref{fig:data-sharing-workflow} shows the practical types of steps that can be taken against the effort required.

Creating good documentation has similar challenges to good writing: it takes time, and it can be hard to know when you are done. Thinking about two features: 1) the impact of data, and 2) the current effort to do each step, provides guidelines for the researcher to decide how much documentation they should provide for their data.

Documentation challenges may evolve over time as the cost of making them easy to prepare and understand change. For example, if new technology automates rigorous data documentation, thorough documentation can take the same time it takes to do poorly now. In this case, there would be no reason why more people cannot do this, even when the benefits are not immediately apparent. So if something might appear to have low impact, it should still be made easy to understand, if it does not take long to prepare that documentation.

\hypertarget{publish-repos}{%
\section{Publishing and repositories}\label{publish-repos}}

It is worth distinguishing between sharing data and publishing data. Data can be shared in numerous ways without going through the trouble of publishing it, which often requires metadata that a human must verify. Data can be shared in numerous ways, including placing it in a repository, packaging it with methods, or using various free tiers of commercial services (such as dropbox or google drive). However, one must publish data when appropriate.

There are three common options for publishing data in research:

\begin{enumerate}
\def\labelenumi{\arabic{enumi}.}
\tightlist
\item
  \textbf{Least Moving Parts}
\item
  \textbf{Domain Specific Venue}
\item
  \textbf{Long Term Archive}
\end{enumerate}

We discuss each of these options and provide recommendations on how to publish data in these areas.

In \textbf{Least Moving Parts}, the data might be published with an R package, or as part of a GitHub release using piggyback (``Piggyback,'' n.d.), or in a serverless database. This approach \textbf{gets the data somewhere rather than nowhere}. Its minimal features means it is simple to maintain. In addition, piggyback also keeps the data in the same location as the code, making it easier and simpler to organize. A downside is that it does not scale to larger data. Self hosting the data is an option, but we discourage this, as it may succumb to bit rot - where data is corrupted, degraded, or servers turn off.

In \textbf{Domain Specific Venue}, data can be published in journal data papers, or venues specific to the type of data. For example, an astronomy project hosts its data at the Sloan Digital Sky Survey (SDSS), and Genetic data can be hosted at GenBank (Benson et al. (2005)).

The purpose, use, and origin of the data is an important component to consider. Data for research has a different domain compared to data collected by governments, or by journalists. Many governments or civil organizations are now making their own data available through a government website interface. Media and journalism groups are also making their data available either through places like GitHub (``Pudding-Data,'' n.d.), or may self host their data, such as five thirty eight (``Data-538,'' n.d.).

This is a good option when the data is appropriate for the domain. This is often decided by community standards. We recommend adhering to community standards for a location to publish your data, as this will likely improve data reuse. The guidelines suggested in this paper for sharing the data should be included.

A \textbf{Long Term Archive} is the best option to share the data. Long term archives provide information such as DOI (Digital Object Identifier) that make it easier to cite. Data can be placed in a long term archive and a DOI can be minted. This DOI can then be used in other venues, such as domain specific or even self hosted, and will ensure that the projects refer back appropriately.

If the dataset you are shipping has a research application, the most relevant home would be the research data repositories \textbf{zenodo}, \textbf{dryad}, or \textbf{Open Science Framework}. Zenodo, Launched in 2013 in a joint collaboration between openAIRE and CERN, provides a free, archival location for any researcher to deposit their datasets. The only limits to file sizes are 50gb for individual files, which is generous enough to accommodate a large number of use cases. Zenodo is able to accommodate larger file sizes upon request. The Dryad Digital Repository ((``Dryad,'' n.d.)) will take data from any field of research, and perform human quality control and assistance of the data, with the ability to link data with a journal publication, in exchange for a data publishing fee. The OSF is a tool that captures the research process online. This ranges from conceiving research ideas, study design, writing papers, to storing research data. The entire history of the project is recorded, so it promotes centralized, and transparent workflows. Although not specifically designed for data, like Dryad or Zenodo, OSF does provide a DOI minting service, and a more wholistic approach, which might be appealing to users (Spies et al. 2012).

\hypertarget{publishing-data-in-a-data-repository}{%
\subsection{Publishing data in a data repository}\label{publishing-data-in-a-data-repository}}

Scientists have numerous venues to deposit their data, many of which are ephemeral, despite the convenience they offer. For data that are not narrow in scope with limited potential for reuse, we recommend publishing data in an established data repository such as Dryad. In addition to providing a long-term home for data, these repositories also include data curation by a professional data manager, compliance with open source standards, and metrics.

\hypertarget{publishing-data-through-a-data-journal}{%
\subsection{Publishing data through a data journal}\label{publishing-data-through-a-data-journal}}

Data used in publications are often shared in the supplementary materials of articles, or served on repositories such as the Dryad Digital Repository ((``Dryad,'' n.d.)). Dryad makes data from scientific publications discoverable, reusable, and citable. It is well funded through grants from the National Science Foundation and European Commission.

To provide better context around the data used in research and better expose data for reuse, journals are now adding ``data papers''. These are specifically designed for publishing articles about the data, and sharing it. This benefits both researchers and readers. Researchers receive credit for data they have collected or produced, and readers get more context about the data.

Data papers are similar to research articles, they have titles, authors, affiliations, abstract, and references. They generally require an explanation of why the data is useful to others, a direct link to the data, description of the design, materials, and methods. Other information on the subject area, data type, format, and related articles are usually required.

Whilst useful, these requirements do not tell the author how to actually structure the data and folders for reuse. Instead, they provide ideas on what they should include. This is a useful step towards improving data reuse, but it \emph{lacks some minimal structure that allows a researcher to have a predictable way to access and interpret the data}.

Other journals operating in this space include journals like ``data in brief'', ``Data'', and ``Nature Scientific Data''. Guidelines for what is required in the content of the paper, and the required information along with the data (meta data, data format, etc.) vary by journal.

\hypertarget{tooling}{%
\section{Tooling for packaging data}\label{tooling}}

Tooling for producing data documentation information speeds up and simplifies the process of sharing data. Machine-readable metadata that can be indexed by google is created using the \texttt{dataspice} package (\emph{Dataspice: Create Lightweight Schema.org Descriptions of Dataset} 2018). To help create a data dictionary, we recommend the \texttt{codebook} package in R (Arslan 2019), which also generates machine readable metadata. Data dictionaries are implemented in other software such as STATA, which provides a ``codebook'' command. Data can be packaged up in a ``data package'' with DataPackageR (Greg Finak 2019), which provides tools to wrap up data into an R package, while providing helpers with MD5sum checks that allow checksum file comparisons for versioning. Note that is different to Frictionless Data's tabular data package spec (Fowler, Barratt, and Walsh 2017).

\hypertarget{example-data}{%
\section{Example datasets}\label{example-data}}

We now explore the variety of documentation practices of a few selected datasets.

\hypertarget{ex-data-1}{%
\subsection{Dataset 1: Forest Census Plot on Barro Colorado Island}\label{ex-data-1}}

Long-term field surveys, where the same type of data is repeatedly measured and collected over a long period of time, are an example of data collection where the time and financial investment would necessitate meticulously curated and easily accessible metadata. Both from the fact that the same protocol is being followed year after year, and that field data collection efforts are quite expensive, these data need to be well documented, with documentation available alongside the data.

One example of a very laborious, long-term study is the 50-hectare plot on Barro Colorado Island in Panama. Researchers at the field station have censused every single tree in a pre-defined plot 7 times over the past 30 years. More than a million trees have been counted. The data and metadata however are unnecessarily difficult to reach. To obtain the data, one must fill out a form and agree to terms and conditions which require approval and authorship prior to publication. The biggest challenge with this study is that the data are stored on a personal FTP server of one of the authors. While the data are available in CSV and binary \texttt{Rdata} formats, the data storage servers do not have any metadata, README files or a data dictionary. A separate, public archive hosts the metadata (\url{https://repository.si.edu/handle/10088/20925}) in a PDF file (\url{https://repository.si.edu/bitstream/handle/10088/20925/RoutputFull.pdf?sequence=1\&isAllowed=y}) that describe the fields.

\hypertarget{sdss-data}{%
\subsection{Dataset 2: Sensor data}\label{sdss-data}}

Datasets obtained from sensors such as meteorological data are typically easy to prepare and understand. This is because sensors measure specific features, so the description of data type happens upstream at the instrument-design level, and flows down to data collection. The telescope data from the Sloan Digital Sky Survey (SDSS) is a great example of sensor data.

The SDSS data includes photometric and spectroscopic information obtained from a 2.5m telescope at Apache Point Observatory in New Mexico, operating since 2000, producing over 200Gb of data every day (York 2000; Blanton et al. 2017; ``Sdss-Website,'' n.d.). This data has been used to produce the most detailed, three dimensional maps of the universe ever made.

\begin{figure}

{\centering \includegraphics[width=0.6\linewidth]{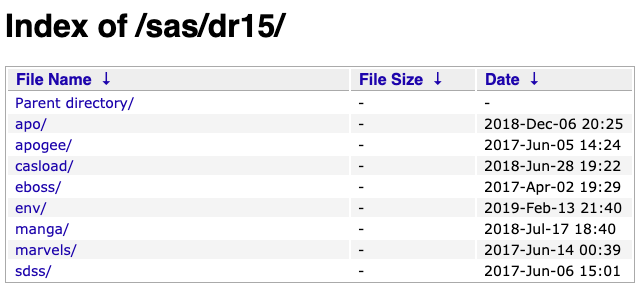} 

}

\caption{Screenshot of Raw data avaialbe through DR15 FITS.}\label{fig:sdss-raw-data-screenshot}
\end{figure}

The data are free to use and publicly accessible, with the interface to the data being designed to cover a wide range of skills. For example, the marvin service to streamline access and visualisation of MaNGA data (Cherinka et al. 2018), through to raw data access (see Figure \ref{fig:sdss-raw-data-screenshot}).

The telescope data is very high quality, with the following features:

\begin{itemize}
\tightlist
\item
  Metadata in the machine readable standard, SCHEMA
\item
  Data in multiple formats (e.g., .csv, .fits, and database)
\item
  Previously released data available in archives
\item
  Entire datasets available for download
\item
  Raw data before processed into database is also available
\end{itemize}

\begin{figure}

{\centering \includegraphics[width=0.9\linewidth]{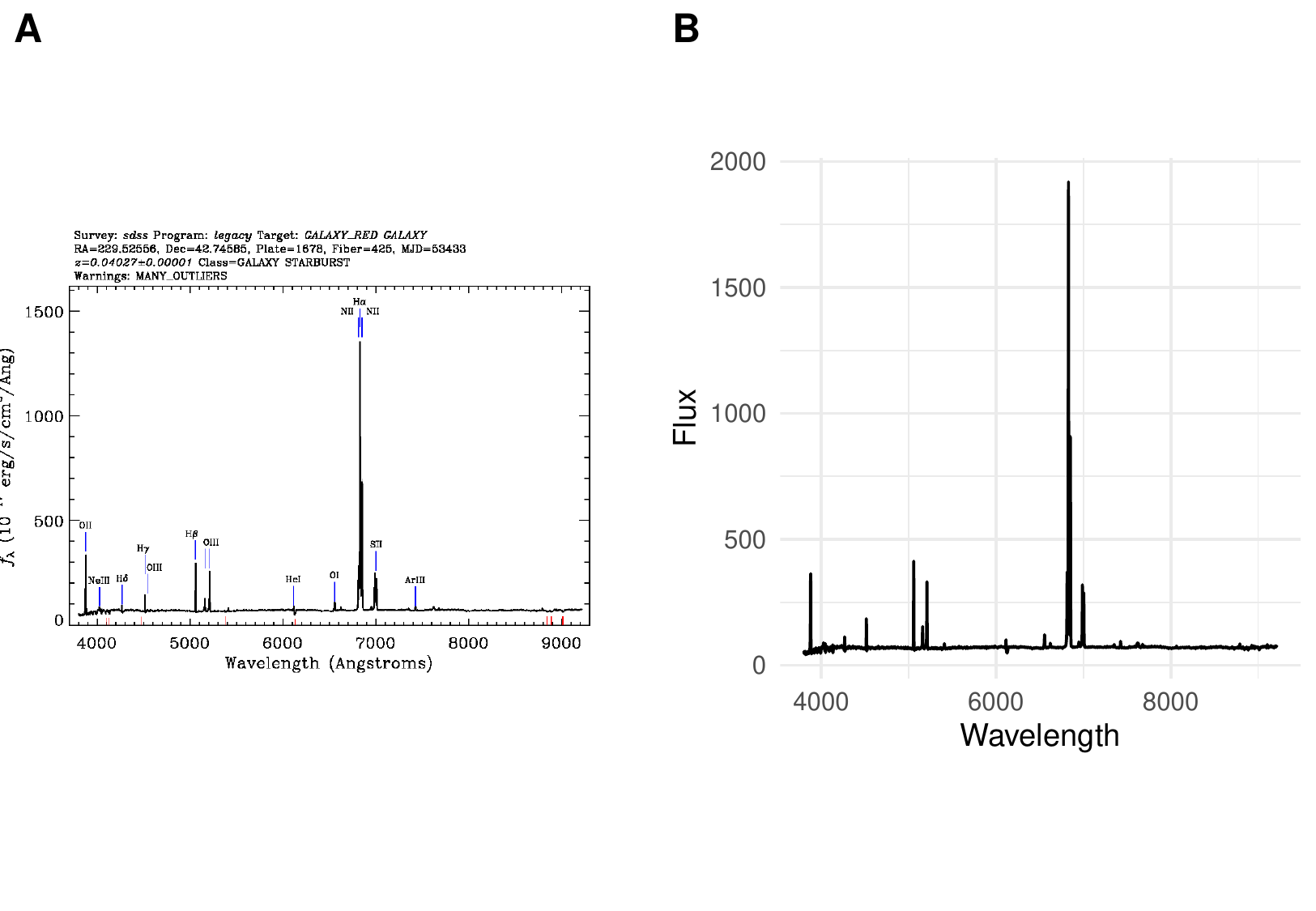} 

}

\caption{Visualisations from SDSS can be effecticely replicated. The spectra image of wavelength against flux from SDSS from two sources. Part A shows the example image taken from SDSS website. Part B shows the replicated figure using data provided as CSV - although there is some data cleaning or filtering since the numbers are not identical, it still demonstrates the great potential of reproducibility.}\label{fig:sdss-spectra}
\end{figure}

For example, the optical and spectral data can be accessed in FITS data format, or even as a plain CSV, the first few rows shown in Table \ref{tab:spectra-tab}. Wavelength plotted against Flux is shown in Figure \ref{fig:sdss-spectra}, with \ref{fig:sdss-spectra}A showing the output from the SDSS website, and Figure \ref{fig:sdss-spectra}B showing the output from the CSV. The fact that figure Figure \ref{fig:sdss-spectra}A is virtually replicated in Figure \ref{fig:sdss-spectra}B demonstrates great reproducibility.

\begin{table}[!h]

\caption{\label{tab:spectra-tab}Example spectra data from SDSS, showing values for Wavelength, Flux, BestFit, and SkyFlux.}
\centering
\begin{tabular}[t]{rrrr}
\toprule
Wavelength & Flux & BestFit & SkyFlux\\
\midrule
\rowcolor{gray!6}  3801.893 & 48.513 & 44.655 & 14.002\\
3802.770 & 54.516 & 42.890 & 14.113\\
\rowcolor{gray!6}  3803.645 & 52.393 & 47.813 & 14.142\\
3804.522 & 45.273 & 42.612 & 14.186\\
\rowcolor{gray!6}  3805.397 & 51.529 & 39.659 & 14.221\\
\addlinespace
3806.273 & 44.530 & 44.183 & 14.349\\
\bottomrule
\end{tabular}
\end{table}

The SDSS provides data at different levels of complexity for different degrees of understanding. It is a huge project involving many people, and a staggering effort to prepare. Despite this, it is still very easy to understand. This is a further reflection on the idea that high effort can create highly understandable data, (see Figure \ref{fig:effort-understanding}). The impact of this data is high, having changed the way we understand the past, present and future of the universe itself. This impact is surely due to the care and thought put into making the data accessible and public. It is a statement of what is possible with good data sharing practices.

\hypertarget{ex-data-3}{%
\subsection{Dataset 3: Most other datasets}\label{ex-data-3}}

The vast majority of datasets are handled in an ad-hoc manner (Michener et al. 2001) and dumped in transient data stores with little usable metadata. The value of these datasets decline over time until they are useless. We recommend that researchers consider the tradeoff in long-term value of the dataset and effort required to document and archive.

\hypertarget{ten-rules}{%
\section{Ten simple rules for publishing data}\label{ten-rules}}

\begin{enumerate}
\def\labelenumi{\arabic{enumi}.}
\item
  \textbf{Decide whether publishing is appropriate for your data}
  It is critical to publish your data if they are the basis for a peer-reviewed research publication. To be broadly useful, your data must be deposited in a permanent archive, to ensure that it does not disappear from a ephemeral location such as your university website. It should also contain useful metadata so that someone outside the project team can understand what the data, and variables, mean. However, this level of effort is not always critical for all data science efforts. For small-scale projects such as ones where one might generate simulated datasets, this level of curation is highly unnecessary. Here making data available in a transient location like GitHub or a website is sufficient.
\item
  \textbf{Include a README} file with your data archive. README files have a long history in software (\url{https://medium.com/@NSomar/readme-md-history-and-components-a365aff07f10}) and are named so that the ASCII systems capital letters filename would show this file first, making it the obvious place to put relevant information. In the context of datasets, READMEs are particularly useful when there are no reliable standards. The best practice here is to have one README per dataset, with names that easily associate with the corresponding data files. In addition to metadata about the data such as variable names, units of measurement and code books, the README should also contain information on licenses, authors/copyright, a title, and dates and locations where the data were collected. Additionally keywords, author identifiers such as ORCID, and funder information would be useful.
\item
  \textbf{Provide a data dictionary} or glossary for each variable. Data dictionaries provide a free-form way to document the data, their structure and underlying details of how they were collected (\url{https://www.emgo.nl/kc/codebook/}). More importantly they provide a way for future consumers of the data to understand how the variables were named, how missing data were documented, along with any additional information that might impact how the data are interpreted.
\item
  \textbf{Provide a machine readable format} for your data. When possible, provide machine readable metadata that map on to the open standards of \href{https://schema.org/}{schema.org} and \href{https://en.wikipedia.org/wiki/JSON-LD}{JSON-LD}. These metadata provide all of the information described in the README best practices (rule 2), but in a machine readable way and include much of the same information such as name, description, spatial/temporal coverage etc. One way to test if your metadata are machine readable is to use Google's structured testing data (\url{https://search.google.com/structured-data/testing-tool/u/0/}) to verify the integrity of the metadata.
\item
  \textbf{Provide the data in its rawest form} in a folder called ``data-raw''. Keeping your data raw (also sometimes referred to as the sushi principle) is the safest way to ensure that your analysis can be reproduced from scratch. This approach not only lets you trace the provenance of any analysis but it also ensures further use of the data that a derived dataset may prevent.
\item
  \textbf{Provide {[}open source?{]} scripts} used to clean data from rawest form into tidy/analysis ready format. Raw data is often unusable without further processing (data munging or data cleaning), which are the steps necessary to detect and clean inaccurate, incomplete records, and missing values from a dataset. While datasets can be cleaned interactively, this approach is often very difficult to reproduce. It is a better practice to use scripts to batch process data cleaning. Scripts can be verified, version controlled and rerun without much overhead when mistakes are uncovered. These scripts describe the steps taken to prepare the data, which helps explain and document any decisions taken during the cleaning phase. These should ideally operate on some raw data stored in the ``data-raw'' folder (rule 5).
\item
  \textbf{Keep additional copies in more accessible locations}: Even if you archive the data somewhere long-term, keep a second (or additional copies) in locations that are more readily and easily accessible for data analysis workflows. These include services like GitHub, GitHub LFS and others where fast content delivery networks (CDNs) make access and reuse practical. In the event that these services shut down or become unavailable, a long-term archival copy can still be accessed from a permanent repository such as Zenodo or Dryad and populated into a new data hosting service or technology.
\item
  \textbf{Use a hash function like MD5 checksum} to ensure that the data you deposited are the same that are being made available to others. Hash values such as MD5 are short numeric values that can serve as digital signatures for large amounts of data. By sharing the hash in a README, authors can ensure that the data, particularly the version of data being used by the reader is the same.
\item
  \textbf{Only add a citation if your data has a DOI}. A citation only makes sense when your data has a DataCite compliant DOI, which is automatically provided when data is published in repositories like Zenodo and Dryad. While a citation may not accrue references, without a DOI it is guaranteed not to.
\item
  \textbf{Stick with simple data formats} to ensure long-term usefulness of datasets in a way that is not tied to transient technologies (such as ever changing spreadsheet standards), store the data as plain text (e.g., CSV) where possible. This can take the form of comma, or tab separated files in most cases. This will ensure transparency, and future proof your data.
\end{enumerate}

\hypertarget{conclusions}{%
\section{Conclusions}\label{conclusions}}

The open science literature has been advocating the importance of data sharing for a long time. Many of these articles appeal to the broader benefits of data sharing, but rarely talk about the practical considerations of making data available alongside code. Trivial reproducibility has always relied upon the idea that as long as code exists on a platform (e.g., GitHub), and clearly lists open source dependencies, one should be able to run the code at a future time. Containerization efforts such as Docker have made it even easier to capture system dependencies, and rapidly launch containers. But true reproducibility requires not just code, but also data.

Over the years researchers have made data available in a multitude of unreliable ways, including offering downloads from university websites, FTP repositories, and password protected servers, all of which are prone to bit rot. In rare cases where the data are deposited in a permanent archive, insufficient metadata and missing data processing scripts have made it harder and more time intensive to map these to code. Our goal here is to describe a variety of ways a researcher can ship data, analogous to how easy it has become to push code to services like GitHub. The choice of technology and method all depends on the nature of the data, and its size and complexity. Once the issue of access is solved, we also discuss how much metadata to include in order to make the data useful. Ultimately there is a tradeoff to consider on a case by case basis. Long-term studies with complex protocols require structured metadata, while more standardized datasets such as those coming from sensors require far less effort.

The key message for the reader is that accessing data for reproducibility should come with minimal barriers. Having users jump through data usage forms or other barriers is enough of a roadblock to dissuade users, which over time will make it hard to track the provenance of critical studies. For data with broad impact, one should further invest effort in documenting the metadata (either unstructured or structured depending on the complexity) and also focus on ensuring that at least one archival copy exists in a permanent archive. Following this continuum can ensure that more and more computational work becomes readily reproducible without unreasonable effort.

\hypertarget{acknowledgements}{%
\section{Acknowledgements}\label{acknowledgements}}

\begin{itemize}
\tightlist
\item
  Mile McBain
\item
  Anna Krystalli
\item
  Daniella Lowenberg
\item
  ACEMS International Mobility Programme
\item
  Helmsley Charitable Trust, Gordon and Betty Moore Foundation, Sloan Foundation.
\end{itemize}

\hypertarget{literature-cited}{%
\section*{Literature Cited}\label{literature-cited}}
\addcontentsline{toc}{section}{Literature Cited}

\hypertarget{refs}{}
\begin{cslreferences}
\leavevmode\hypertarget{ref-Arslan2019}{}%
Arslan, Ruben C. 2019. ``How to Automatically Document Data with the Codebook Package to Facilitate Data Reuse.'' \emph{Advances in Methods and Practices in Psychological Science} 2 (2): 169--87.

\leavevmode\hypertarget{ref-Barnes2010}{}%
Barnes, Nick. 2010. ``Publish Your Computer Code: It Is Good Enough.'' \emph{Nature} 467 (7317): 753--53. \url{https://doi.org/10.1038/467753a}.

\leavevmode\hypertarget{ref-GenBank}{}%
Benson, Dennis A, Ilene Karsch-Mizrachi, David J Lipman, James Ostell, and David L Wheeler. 2005. ``GenBank.'' \emph{Nucleic Acids Research} 33 (Database issue): D34--8.

\leavevmode\hypertarget{ref-julia}{}%
Bezanson, Jeff, Alan Edelman, Stefan Karpinski, and Viral B Shah. 2017. ``Julia: A Fresh Approach to Numerical Computing.'' \emph{SIAM Review} 59 (1): 65--98. \url{https://doi.org/10.1137/141000671}.

\leavevmode\hypertarget{ref-sdss-four-paper}{}%
Blanton, Michael R, Matthew A Bershady, Bela Abolfathi, Franco D Albareti, Carlos Allende Prieto, Andres Almeida, Javier Alonso-García, et al. 2017. ``Sloan Digital Sky Survey IV: Mapping the Milky Way, Nearby Galaxies, and the Distant Universe,'' February. \url{http://arxiv.org/abs/1703.00052}.

\leavevmode\hypertarget{ref-arkdb}{}%
Boettiger, Carl. 2018. \emph{Arkdb: Archive and Unarchive Databases Using Flat Files}. \url{https://CRAN.R-project.org/package=arkdb}.

\leavevmode\hypertarget{ref-Boettiger2015}{}%
---------. 2015. ``An Introduction to Docker for Reproducible Research.'' \emph{ACM SIGOPS Operating Systems Review} 49 (1): 71--79. \url{https://doi.org/10.1145/2723872.2723882}.

\leavevmode\hypertarget{ref-Broman2017}{}%
Broman, Karl W., and Kara H. Woo. 2017. ``Data Organization in Spreadsheets.'' \emph{The American Statistician} 72 (1): 2--10. \url{https://doi.org/10.1080/00031305.2017.1375989}.

\leavevmode\hypertarget{ref-Brooke_Anderson2017}{}%
Brooke Anderson, G, and Dirk Eddelbuettel. 2017. ``Hosting Data Packages via Drat: A Case Study with Hurricane Exposure Data.'' \emph{The R Journal} 9 (1): 486--97.

\leavevmode\hypertarget{ref-google-data-search}{}%
Castelvecchi, Davide. 2018. ``Google Unveils Search Engine for Open Data.'' \emph{Nature} 561 (7722): 161--62.

\leavevmode\hypertarget{ref-cc0-long}{}%
``CC0 Full License.'' n.d. https://creativecommons.org/publicdomain/zero/1.0/legalcode.

\leavevmode\hypertarget{ref-cc0-short}{}%
``CC0 Short Guide.'' n.d. https://creativecommons.org/publicdomain/zero/1.0/.

\leavevmode\hypertarget{ref-ccby-long}{}%
``CCBY Full License.'' n.d. https://creativecommons.org/licenses/by/4.0/legalcode.

\leavevmode\hypertarget{ref-ccby-short}{}%
``CCBY Short Guide.'' n.d. https://creativecommons.org/licenses/by/4.0/.

\leavevmode\hypertarget{ref-marvin-service}{}%
Cherinka, Brian, Brett H Andrews, José Sánchez-Gallego, Joel Brownstein, María Argudo-Fernández, Michael Blanton, Kevin Bundy, et al. 2018. ``Marvin: A Toolkit for Streamlined Access and Visualization of the SDSS-IV MaNGA Data Set,'' December. \url{http://arxiv.org/abs/1812.03833}.

\leavevmode\hypertarget{ref-choose-cc0}{}%
``Choose-Cc0.'' n.d. https://creativecommons.org/choose/zero/.

\leavevmode\hypertarget{ref-McGill-codebook}{}%
``Codebook Cookbook: A Guide to Writing a Good Codebook for Data Analysis Projects in Medicine.'' n.d. Accessed December 18, 2018. \url{http://www.medicine.mcgill.ca/epidemiology/joseph/pbelisle/CodebookCookbook/CodebookCookbook.pdf}.

\leavevmode\hypertarget{ref-cran}{}%
``Cran.'' n.d. https://cran.r-project.org/.

\leavevmode\hypertarget{ref-csvy}{}%
``Csvy.'' n.d. http://csvy.org/.

\leavevmode\hypertarget{ref-data-538}{}%
``Data-538.'' n.d. https://data.fivethirtyeight.com/.

\leavevmode\hypertarget{ref-dataspice}{}%
\emph{Dataspice: Create Lightweight Schema.org Descriptions of Dataset}. 2018. \url{https://github.com/ropenscilabs/dataspice}.

\leavevmode\hypertarget{ref-Donoho2017}{}%
Donoho, David. 2017. ``50 Years of Data Science.'' \emph{Journal of Computational and Graphical Statistics: A Joint Publication of American Statistical Association, Institute of Mathematical Statistics, Interface Foundation of North America} 26 (4): 745--66.

\leavevmode\hypertarget{ref-dryad}{}%
``Dryad.'' n.d. https://datadryad.org/.

\leavevmode\hypertarget{ref-Ellis2017}{}%
Ellis, Shannon E., and Jeffrey T. Leek. 2017. ``How to Share Data for Collaboration.'' \emph{The American Statistician} 72 (1): 53--57. \url{https://doi.org/10.1080/00031305.2017.1375987}.

\leavevmode\hypertarget{ref-Fowler2017}{}%
Fowler, Dan, Jo Barratt, and Paul Walsh. 2017. ``Frictionless Data: Making Research Data Quality Visible.'' \emph{IJDC} 12 (2): 274--85.

\leavevmode\hypertarget{ref-Gentleman2004}{}%
Gentleman, Robert C, Vincent J Carey, Douglas M Bates, Ben Bolstad, Marcel Dettling, Sandrine Dudoit, Byron Ellis, et al. 2004. \emph{Genome Biology} 5 (10): R80. \url{https://doi.org/10.1186/gb-2004-5-10-r80}.

\leavevmode\hypertarget{ref-gentleman2007statistical}{}%
Gentleman, Robert, and Duncan Temple Lang. 2007. ``Statistical Analyses and Reproducible Research.'' \emph{Journal of Computational and Graphical Statistics} 16 (1): 1--23.

\leavevmode\hypertarget{ref-git}{}%
``Git.'' n.d. https://git-scm.com/about.

\leavevmode\hypertarget{ref-github}{}%
``Github.'' n.d. https://github.com/.

\leavevmode\hypertarget{ref-DataPackageR}{}%
Greg Finak. 2019. \emph{DataPackageR: Construct Reproducible Analytic Data Sets as R Packages}. \url{https://doi.org/10.5281/zenodo.2620378}.

\leavevmode\hypertarget{ref-julia-pkgman}{}%
``Julia-Pkgman.'' n.d. https://pkg.julialang.org/.

\leavevmode\hypertarget{ref-Jupyter2018}{}%
Jupyter. 2018. ``Binder 2.0 - Reproducible, Interactive, Sharable Environments for Science at Scale.'' \emph{Proceedings of the 17th Python in Science Conference}. \url{https://doi.org/10.25080/Majora-4af1f417-011}.

\leavevmode\hypertarget{ref-Kirk2019}{}%
Kirk, Rebecca, and Larry Norton. 2019. ``Supporting Data Sharing.'' \emph{NPJ Breast Cancer} 5 (February): 8.

\leavevmode\hypertarget{ref-McKiernan2016}{}%
McKiernan, Erin C, Philip E Bourne, C Titus Brown, Stuart Buck, Amye Kenall, Jennifer Lin, Damon McDougall, et al. 2016. ``How Open Science Helps Researchers Succeed.'' \emph{eLife} 5 (July). \url{https://doi.org/10.7554/elife.16800}.

\leavevmode\hypertarget{ref-Michener2001}{}%
Michener, William, James Brunt, Kristin Vanderbilt, and Sevilleta Program. 2001. ``Ecological Informatics: A Long-Term Ecological Research Perspective'' 51 (January).

\leavevmode\hypertarget{ref-Peng2011}{}%
Peng, R. D. 2011. ``Reproducible Research in Computational Science.'' \emph{Science} 334 (6060): 1226--7. \url{https://doi.org/10.1126/science.1213847}.

\leavevmode\hypertarget{ref-piggyback}{}%
``Piggyback.'' n.d. https://github.com/ropensci/piggyback.

\leavevmode\hypertarget{ref-plos-comp-bio-data}{}%
``PLOS Computational Biology.'' n.d. \emph{PLOS ONE}. Public Library of Science. \url{https://journals.plos.org/ploscompbiol/s/data-availability}.

\leavevmode\hypertarget{ref-Poisot2019}{}%
Poisot, Timothée, Anne Bruneau, Andrew Gonzalez, Dominique Gravel, and Pedro Peres-Neto. 2019. ``Ecological Data Should Not Be so Hard to Find and Reuse.'' \emph{Trends in Ecology \& Evolution} 34 (6): 494--96. \url{https://doi.org/10.1016/j.tree.2019.04.005}.

\leavevmode\hypertarget{ref-Popkin2019}{}%
Popkin, Gabriel. 2019. ``Data Sharing and How It Can Benefit Your Scientific Career.'' \emph{Nature} 569 (7756): 445--47.

\leavevmode\hypertarget{ref-pudding-data}{}%
``Pudding-Data.'' n.d. https://github.com/the-pudding/data.

\leavevmode\hypertarget{ref-pypi}{}%
``Pypi.'' n.d. https://pypi.org/.

\leavevmode\hypertarget{ref-python}{}%
``Python.'' n.d. https://www.python.org/.

\leavevmode\hypertarget{ref-Ram2013}{}%
Ram, Karthik. 2013. ``Git Can Facilitate Greater Reproducibility and Increased Transparency in Science.'' \emph{Source Code for Biology and Medicine} 8 (1). \url{https://doi.org/10.1186/1751-0473-8-7}.

\leavevmode\hypertarget{ref-rcore}{}%
R Core Team. 2019. \emph{R: A Language and Environment for Statistical Computing}. Vienna, Austria: R Foundation for Statistical Computing. \url{https://www.R-project.org/}.

\leavevmode\hypertarget{ref-Rowhani-Farid2016}{}%
Rowhani-Farid, Anisa, and Adrian G Barnett. 2016. ``Has Open Data Arrived at the British Medical Journal (BMJ)? An Observational Study.'' \emph{BMJ Open} 6 (10): e011784.

\leavevmode\hypertarget{ref-sdss-website}{}%
``Sdss-Website.'' n.d. https://www.sdss.org/.

\leavevmode\hypertarget{ref-OSF}{}%
Spies, Jeffrey R, Brian A Nosek, Sheila Miguez, Ben B Blohowiak, Michael Cohn, Elizabeth Bartmess, and L Simon. 2012. ``Open Science Framework.'' \emph{Open Science Collaboration}.

\leavevmode\hypertarget{ref-Stodden2018}{}%
Stodden, Victoria, Jennifer Seiler, and Zhaokun Ma. 2018. ``An Empirical Analysis of Journal Policy Effectiveness for Computational Reproducibility.'' \emph{Proceedings of the National Academy of Sciences of the United States of America} 115 (11): 2584--9.

\leavevmode\hypertarget{ref-tidy-tuesday-incarcerate}{}%
``Tidy-Tuesday-Incarcerate.'' n.d. https://github.com/rfordatascience/tidytuesday/tree/master/data/2019/2019-01-22.

\leavevmode\hypertarget{ref-White2013}{}%
White, Ethan P, Elita Baldridge, Zachary T Brym, Kenneth J Locey, Daniel J McGlinn, and Sarah R Supp. 2013. ``Nine Simple Ways to Make It Easier to (Re)use Your Data.'' \emph{Ideas in Ecology and Evolution} 6 (2).

\leavevmode\hypertarget{ref-Wickham2014}{}%
Wickham, Hadley. 2014. ``Tidy Data.'' \emph{Journal of Statistical Software} 59 (10). \url{https://doi.org/10.18637/jss.v059.i10}.

\leavevmode\hypertarget{ref-nycflights}{}%
---------. 2018. \emph{Nycflights13: Flights That Departed Nyc in 2013}. \url{https://CRAN.R-project.org/package=nycflights13}.

\leavevmode\hypertarget{ref-Wilkinson2016}{}%
Wilkinson, Mark D., Michel Dumontier, IJsbrand Jan Aalbersberg, Gabrielle Appleton, Myles Axton, Arie Baak, Niklas Blomberg, et al. 2016. ``The FAIR Guiding Principles for Scientific Data Management and Stewardship.'' \emph{Scientific Data} 3 (March): 160018. \url{https://doi.org/10.1038/sdata.2016.18}.

\leavevmode\hypertarget{ref-sdss-one-paper}{}%
York, D G. 2000. ``The Sloan Digital Sky Survey: Technical Summary,'' June. \url{http://arxiv.org/abs/astro-ph/0006396}.

\leavevmode\hypertarget{ref-zenodo}{}%
Zenodo. 2016. ``Zenodo.''

\leavevmode\hypertarget{ref-EML-about}{}%
n.d. Accessed December 18, 2018. \url{https://knb.ecoinformatics.org/\#external//emlparser/docs/index.html}.
\end{cslreferences}

\end{document}